\documentclass[aps,prb,twocolumn,groupedaddress,showpacs]{revtex4}
\usepackage{graphicx}
\usepackage{helvet}
\usepackage{amsmath,amssymb}
\usepackage{bm}
\usepackage{subfigure}

\makeatletter
\newcommand{\Rmnum}[1]{\expandafter\@slowromancap\romannumeral #1@}
\makeatother
\begin{document}

\title{$^{13}$C NMR Study on the Charge-Disproportionated Conducting State 
in the Quasi-Two-Dimensional Organic Conductor $\alpha$-(BEDT-TTF)$_2$I$_3$}
\author{Michihiro Hirata, Kyohei Ishikawa, Kazuya Miyagawa}
\author{Kazushi Kanoda}
\affiliation{Department of Applied Physics, University of Tokyo, Bunkyo City, Tokyo, 113-8656, Japan}
\author{Masafumi Tamura}
\affiliation{Department of Physics, Faculty of Science and Technology, Tokyo University of Science, Noda, Chiba, 278-8510, Japan} 
 
\date{\today}

\begin{abstract}
	The conducting state of the quasi-two-dimensional organic conductor, $\alpha$-(BEDT-TTF)$_2$I$_3$, at ambient pressure is investigated with $^{13}$C NMR measurements, which separate the local electronic states at three nonequivalent molecular sites (A, B, and C). The spin susceptibility and electron correlation effect are revealed in a locally resolved manner. While there is no remarkable site-dependence around room temperature, the local spin susceptibility gradually disproportionates among the nonequivalent sites with decreasing temperature. The disproportionation-ratio yields 5:4:6 for A:B:C molecules at 140 K. Distinct site- and temperature-dependences are also observed in the Korringa ratio, $\mathcal{K}_i \propto (1/T_{1}T)_iK^{-2}_i$ ($i$ = A, B, and C), which is a measure of the strength and the type of electron correlations. The values of $\mathcal{K}_i$ point to sizable antiferromagnetic spin correlation. We argue the present results in terms of the theoretical prediction of the peculiar site-specific reciprocal-space ($\bm{k}$-space) anisotropy on the tilted Dirac cone, and discuss the $\bm{k}$-dependent profiles of the spin susceptibility and electron correlation on the cone.
\end{abstract}

\maketitle

\section{Introduction}
\label{intro}
\begin{figure}
 \includegraphics[width=8.6cm]{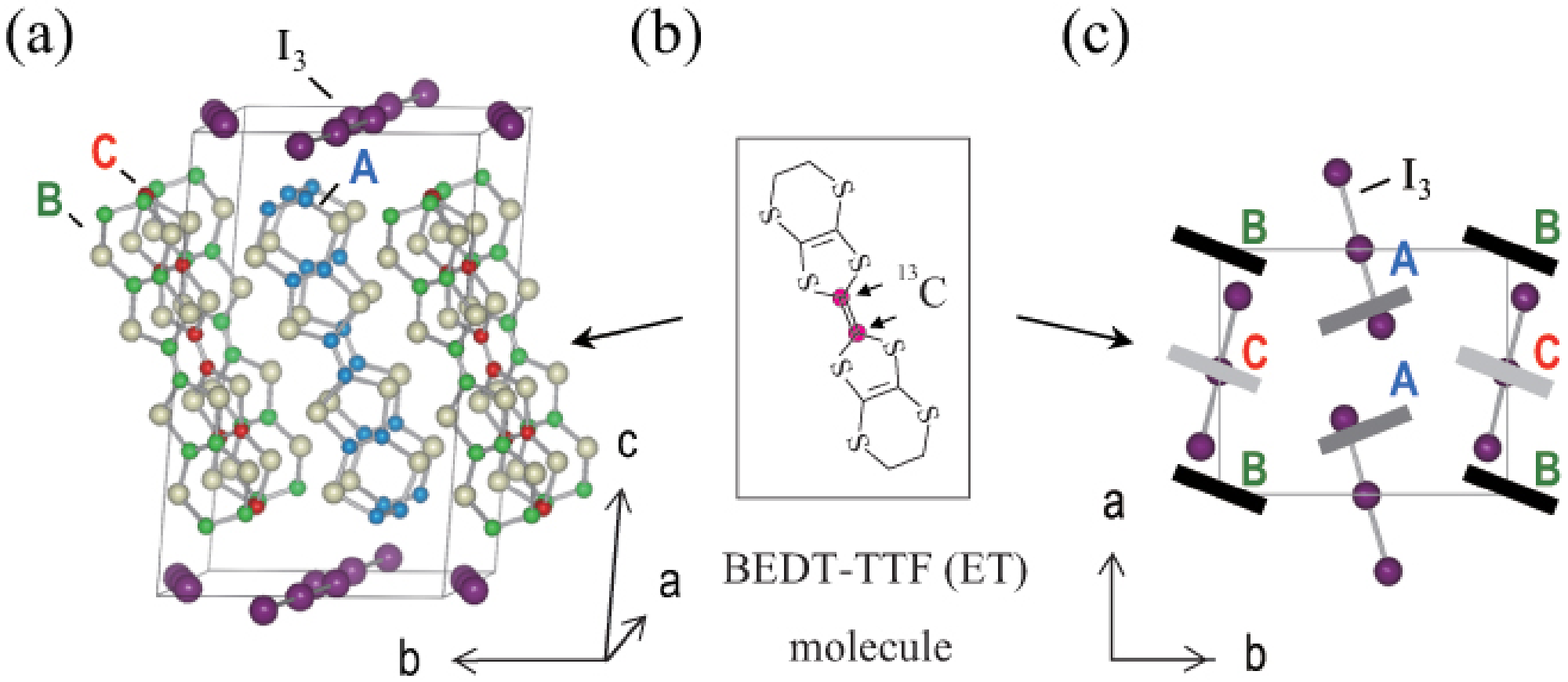}
  \caption{\label{fig1} (Color online) (a) Side view of the crystal structure of $\alpha$-I$_3$. (b) Molecular structure of BEDT-TTF (ET). Arrows indicate the position of carbon atoms enriched by $^{13}$C isotopes with 99\% concentration. (c) Top view of the schematic crystal structure in the conducting layer. Rectangular represents nonequivalent ET molecules (= A, B, and C) in the unit cell.}
\end{figure}

   The organic charge-transfer salt $\alpha$-(BEDT-TTF)$_2$I$_3$ (abbreviated as $\alpha$-I$_3$ hereafter) is a quasi-two-dimensional (Q2D) electron system with a 1/4-filled hole band which exhibits strong electron correlation effects [BEDT-TTF (ET hereafter) is the abbreviation of bisethylenedithio-tetrathiafulvalene; see Fig.~\ref{fig1}(b)].    
   $\alpha$-I$_3$ is comprised of alternately stacked 2D conducting layers of (ET)$^{+1/2}$ molecules and non-magnetic insulating layers of triiodide anions (I$_3$)$^{-1}$ [Fig.~\ref{fig1}(a)].\cite{Bender1} 
   The unit cell contains four ET molecules, three of which are crystallographically nonequivalent [i.e., A, B, and C molecules; see Fig.~\ref{fig1}(c)]. At room temperature, the band structure calculation predicts semi-metallic Fermi surfaces.\cite{Seo2}
   With decreasing temperature, resistivity shows weak temperature dependence down to $T_\textrm{CO} \sim 135$ K,\cite{Bender2} at which the system undergoes a first-order phase transition from a paramagnetic conductor to a nonmagnetic insulator accompanied by an inversion symmetry breaking.\cite{Bender2, Nogami, Kakiuchi, Heidmann, Fortune, Rothaemel, Sugano, Wojciechowski, Kawai, Moldenhauer,Yamamoto}
    Below $T_\textrm{CO}$, electrons are localized on ET lattice sites, forming the horizontal strip type of charge ordering (CO),\cite{Kakiuchi, Wojciechowski, Kawai, Moldenhauer, Yamamoto, Kodama} which is believed to be stabilized by the long-range Coulomb interactions.\cite{Seo1, Tanaka}

   Under hydrostatic pressures, the phase transition tends to be suppressed.\cite{Wojciechowski, Schwenk, Kondo}
   Above $P_\textrm{C} \sim 1.5$ GPa, the charge-ordered state completely disappears and the temperature independent profile of resistivity extends down to several Kelvins. 
   In this pressurized phase, recent band-structure calculations\cite{IshibashiKino, Katayama1, Katayama2} revealed that an unusual electronic structure with the linear dispersion is realized around the Fermi level, $\varepsilon_F$, (i.e., massless Dirac cone) which is partly confirmed by recent magnetotransport experiments.\cite{Tajima1, Osada}  
   In contrast to the graphene monolayer, known as the typical massless Dirac-cone system with isotropic dispersion,\cite{NovoselovKim} the Dirac cone in $\alpha$-I$_3$ is strongly tilted;\cite{Katayama1} the conical slant varies 10 times around the apex in the 2D $\bm{k}$-space.\cite{Katayama2} 
   The tilt stems from the low lattice symmetry in $\alpha$-I$_3$ possessing only inversion symmetry (space group $P$-1). 
   Because of this anisotropy of the band dispersion, a large imbalance in the local electronic density of states around $\varepsilon_F$ is predicted among independent A, B, and C molecules.\cite{Katayama2} 
   Recent $^{13}$C nuclear magnetic resonance ($^{13}$C NMR) experiments under hydrostatic pressures observed large difference in the local spin susceptibilities at these molecular sites.\cite{Hirose, Takano} 
   
   The Dirac-cone picture has been discussed so far only at high hydrostatic pressures ($P > P_\textrm{C}$) in $\alpha$-I$_3$. 
   However, several experiments have recently suggested that it might be also applicable to the low-pressure conducting phase. 
   Resistivity,\cite{Schwenk, Tajima2} Raman,\cite{Wojciechowski} $^{13}$C NMR,\cite{Hirose, Takano} x-ray,\cite{Kondo} Hall,\cite{Tajima2, Tajima3} and thermopower\cite{Bender2,Tajima3} measurements indicate that the ambient- and high-pressure conducting phases have many properties in common.
   Moreover, the band structure calculation\cite{Kondo} suggests that the conical apex (Dirac point) locates at $\sim$ 8 meV below $\varepsilon_F$ even at ambient pressure,\cite{Kondo, IshibashiKino} and can be pushed up to $\varepsilon_F$ by introducing a site-dependent potential imbalance between A, B, and C nonequivalent molecules.\cite{Kondo, Hotta, Mori}
   Aside from this, the ambient-pressure conducting phase exhibits several anomalous properties such as non-Drude optical conductivity just above $T_\textrm{CO}$,\cite{Dressel} and a large charge density disproportionation among A, B, and C molecular sites as observed in x-ray,\cite{Kakiuchi} Raman,\cite{Wojciechowski} and $^{13}$C NMR\cite{Kawai} experiments. 
   On the whole, these findings probably indicate that the conducting state in $\alpha$-I$_3$ is closely associated with the Dirac cone and/or CO fluctuations. 
   However, the local electronic properties are not yet fully understood quantitatively, in particular, the electron correlation effects. 
   
   The purpose of the present study is to clarify the electronic nature in the ambient-pressure conducting state of $\alpha$-I$_3$ in microscopic and site-specific manners. 
   Utilizing $^{13}$C NMR experiments, we explored the local spin susceptibility and electron correlation effects at each nonequivalent molecule ($i$ = A, B, and C), since NMR is capable of probing the local electronic states at these independent molecules separately. 
   As the hyperfine coupling of a nuclear spin with its surrounding electrons is highly anisotropic at the $^{13}$C site, one should know the field-angulear-dependent characteristics of NMR spectra to make a reliable assignment of the NMR line and determine hyperfine tensors, which correspondingly provides a quantitative basis for deducing the local spin susceptibility and the electron correlation effects. 
   Thus, we measured the angular dependences of the NMR Knight shift, $K_i$, with rotating the external field in the planes parallel and perpendicular to the conducting layers [i.e., the $ab$ plane in Fig.~\ref{fig1}(c)] at each temperature measured. 
   The $^{13}$C hyperfine-shift tensors and the local spin susceptibility were determined from the angular-dependent profile of $K_i$ at each independent molecule. 
   The nuclear spin-lattice relaxation rate, $(1/T_1)_i$, was measured under the in-plane fields in order to evaluate the Korringa ratio, $\mathcal{K}_i \propto (1/T_{1}T)_iK^{-2}_i$, which represents the strength and type of the electron correlations.
   The results are discussed in connection with the effect of CO fluctuations and Dirac cone.

\section{Methods}
   To evaluate the local electronic properties in the ambient-pressure paramagnetic conducting state of $\alpha$-I$_3$, we performed $^{13}$C NMR measurements at external field $H$ of 6.00 T applied in the $ab$ and $bc$ planes in the temperature range from room temperature down to 140 K ($> T_\textrm{CO}$). 
   We grew the single crystal of $\alpha$-I$_3$ of the size $2.5 \times 0.5 \times 0.1$ mm$^{3}$ by the conventional electrochemical oxidization method.
   The central double-bonded carbon atoms in ET were selectively enriched with $^{13}$C isotopes (nuclear spin $I = 1/2$, $\gamma_\textrm{n}/2 \pi$ = 10.705 MHz/T) at 99\% concentration [see Fig.~\ref{fig1}(b)], which means that the observed $^{13}$C NMR spectra come from the central carbon sites.
   The $^{13}$C NMR spectra were obtained by the fast Fourier transformation of the spin echo signals. 
   As the origin of the NMR shift, we referred to the resonance frequency of the $^{13}$C NMR signal of TMS [tetramethylsilane, (CH$_{3}$)$_{4}$Si]. 
   The spin-lattice relaxation rate, $1/T_1$, was obtained by the standard saturation and recovery method.

\section{Results}

\subsection{Angular dependence of the $^{13}$C NMR line shapes and line assignment}
\label{3a}
   The assignment of the observed $^{13}$C lines in the NMR spectra to A, B, and C nonequivalent molecular sites is the starting point of the present work. 
   This can be accomplished by considering the crystal structure, the symmetry of A, B, and C molecules in the crystal, and the nuclear dipolar splitting in the analysis of the angular dependence of $^{13}$C NMR spectra. 
   
   In ET-based organic conductors, all crystallographically nonequivalent molecules contribute to distinct $^{13}$C lines in the $^{13}$C NMR line shapes. 
   In the conducting state of $\alpha$-I$_3$, at least three peaks are thus expected associated with the nonequivalent A, B, and C molecules [Fig.~\ref{fig1}(c)].
   However, each molecule has two neighboring $^{13}$C nuclei [Fig.~\ref{fig1}(b)], which are coupled by the nuclear dipole interaction. 
   This coupling makes a single line split into a doublet or a quartet depending on whether the molecule has inversion symmetry or not, respectively.\cite{Slichter, Kawamoto1}
   The Pake doublet structure should be observed when the neighboring nuclei are equivalent, which is the case in molecules B and C. 
   The Pake doublet is characterized by the nuclear dipolar splitting width (kHz), $d$, given by  
\begin{equation}
d = \frac{3}{2} \frac{\gamma_\textrm{n}^2 \hbar}{r^3}(1-3\sin^2 \zeta_i \cos^2 \eta_i ),
\end{equation}  
where $\gamma_\textrm{n}$ is the gyromagnetic ratio of $^{13}$C nucleus, 2$\pi \hbar$ is Planck's constant, $r$ = 1.360 {\AA} is the distance between two $^{13}$C nuclei,\cite{Sawa} and ($\zeta_i$, $\eta_i$) are the angles formed between the direction of $H$ and the $^{13}$C=$^{13}$C vector of the $i$-site molecule ($i$ = A, B, and C), as illustrated in Fig.~\ref{fig3}(b). 
   The quartet structure is expected when the neighboring $^{13}$C sites are nonequivalent as in molecules A. 
   The quartet is composed of four lines ($\nu = 1 - 4$),\cite{Kawamoto1,Kawamoto2} the line shift $\delta_\nu$ of which is dependent on the nuclear dipolar splitting width $d$ and the NMR shift difference between the two neighboring $^{13}$C sites, $\Delta \delta$; i.e., $\delta_\nu = \delta \pm d/3H \pm \sqrt{(d/3H)^2 + (\Delta \delta)^2} / 2$, where $\delta$ is the averaged shift of the quartet, and equals to the average of the NMR shift at the neighboring $^{13}$C sites in ET without the nuclear dipole coupling. 
   Thus, in the present case, eight $^{13}$C lines composed of two Pake doublets and one quartet are expected in the $^{13}$C NMR spectra.
   
\begin{figure*}
 \includegraphics[width=14cm]{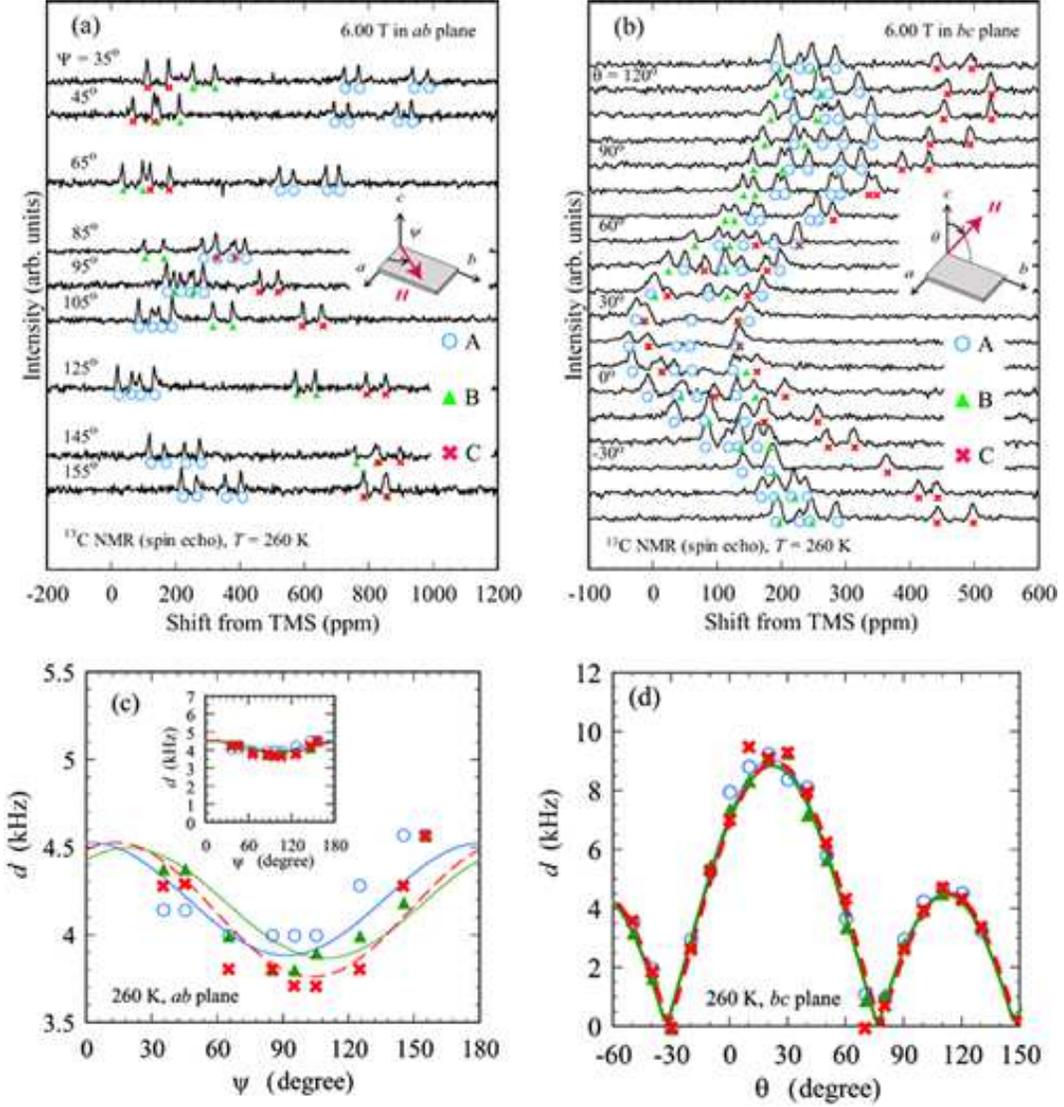}
  \caption{\label{fig2} (Color online) (a), (b) Angular dependences of the $^{13}$C NMR line shapes under (a) $H//ab$ and (b) $H//bc$ at 260 K. The definitions of angles $\psi$ and $\theta$ are given in the insets. [Note that the origins of $\psi$ and $\theta$ are set as $H//b$ and $H//c$, correspondingly]. (c), (d) Nuclear dipolar splitting width $d$ at A, B, and C nonequivalent molecules extracted from Figs.~\ref{fig2}(a) and ~\ref{fig2}(b), respectively. Calculated angular dependences are shown together based on Ref.~\onlinecite{Sawa} (curves). In all figures, symbols represents the data relevant to molecules A (open circles), B (triangles) and C (crosses), respectively.}
\end{figure*}

\begin{figure}
 \includegraphics[width=8.6cm]{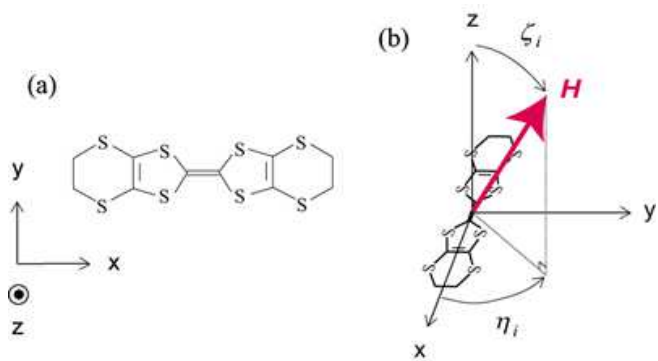} 
  \caption{\label{fig3}(Color online) (a) Molecular principal axes ($x$, $y$, $z$) of ET molecule. (b) Spherical polar coordinate ($\zeta_i$, $\eta_i$) of the external field vector $H$ at $i$-site ET molecule ($i$ = A, B, and C) in the principal axes ($x$, $y$, $z$).\cite{Sawa}} 
\end{figure}

   In order to assign the NMR lines properly, we measured the magnetic-field angular dependences of the NMR spectra.
   Figure~\ref{fig2} presents the typical data at 260 K with the external field $H$ applied within the $ab$ plane ($H//ab$) and $bc$ plane ($H//bc$). 
   The line shapes are strongly dependent on the direction of $H$.
   In Fig.~\ref{fig2}(a), the $^{13}$C lines can be divided into two groups, exhibiting an anti-phase angular dependence as the direction of $H$, $\psi$, is changed within the $ab$ plane [$\psi = 0^\circ$ is set as $H//a$; see the inset of Fig.~\ref{fig2}(a)]. 
   In Fig.~\ref{fig2}(b), on the other hand, all the $^{13}$C lines exhibit an in-phase variation when $H$ is rotated in the $bc$ plane [$\theta = 0^\circ$ is set as $H//c$; see the inset of Fig.~\ref{fig2}(b)].
   These phase relations can be adequately understood from the crystal structure;\cite{Sawa} i.e., the direction of $^{13}$C-2$p_z$ orbitals within the unit cell.
   It is well known that the 2$p_z$ electrons of $^{13}$C atoms produce a dipole-type hyperfine field at the $^{13}$C nuclear position, and give the largest contribution to the $^{13}$C NMR shift.\cite{Kawamoto1, Kawamoto2}
   As schematically shown in Fig.~\ref{fig1}(c), molecules B and C, and molecules A form respectively a stack along the $a$ axis. 
   These stacks build a fishbone-like molecular arrangement in the $ab$ plane, and are approximately connected with mirror operation [($a$, $b$, $c$) $\to$ ($a$, $-b$, $c$)].  
   The 2$p_z$ orbitals, pointing perpendicular to the molecular plane [along the $z$ axis in Fig.~\ref{fig3}(a)], are parallel within the same stack, while they are mutually perpendicular between different stacks. 
   This geometry gives rise to an anti-phase relation in the angular dependence of the NMR shift when the external field is rotated in the $ab$ plane. 
   On the other hand, all the 2$p_z$ orbitals are roughly confined to the $ab$ plane, which is expected to cause an in-phase dependence of the NMR lines when the filed is rotated within the $bc$ plane.

   Referring to the considerations given above, we first assigned the $^{13}$C lines in Fig.~\ref{fig2}(a). 
   The four lines denoted by open circles are attributable to the quartet (molecules A), and the rest four lines to two pairs of the Pake doublets (molecules B and C), as shown in Fig.~\ref{fig2}(a). 
   It is easy to distinguish molecules B and C when one focuses on the dihedral angles between molecules A and B ($\sim 138^\circ$), and A and C ($\sim 131^\circ$).\cite{Sawa}
   Since the former is larger than the latter, molecules B should exhibit larger phase difference from molecules A than molecules C will in the angular dependence of the line shapes. 
   We thus assigned the $^{13}$C lines denoted by triangles and crosses to molecules B and C, respectively, as shown in Fig.~\ref{fig2}(a). 
   The results are consistent with the previous work.\cite{Kawai}
   In Fig.~\ref{fig2}(c), the observed angular dependences of the nuclear dipolar splitting width $d$ for A, B, and C molecules are presented with the corresponding symbols. 
   The solid, dotted, and dashed lines are calculated angular dependences of $d$ for A, B, and C molecules, respectively, with $r$ = 1.360 {\AA} and the molecular orientations obtained from Ref.~\onlinecite{Sawa}. 
   The data agree well with the calculations.  
   
   Next, we focus on Fig.~\ref{fig2}(b). 
   By comparing the NMR spectra at $\theta = \psi = 90^\circ$ in Figs.~\ref{fig2}(a) and ~\ref{fig2}(b) (i.e., $H//b$), the $^{13}$C lines under $H//bc$ were assigned as shown in Fig.~\ref{fig2}(b) [with the same symbols as in Fig.~\ref{fig2}(a)]. 
   Figure~\ref{fig2}(d) presents the angular dependences of the dipolar splitting width $d$ under $H//bc$. 
   Calculated angular dependences are presented together, which properly reproduce the experimental results. 
   All of these analyses thus assure our assignments comprehensively.

\subsection{Angular dependence of the $^{13}$C NMR shift and determination of the hyperfine-shift tensors}
\label{3b}
\begin{figure}
\includegraphics[width=8.6cm]{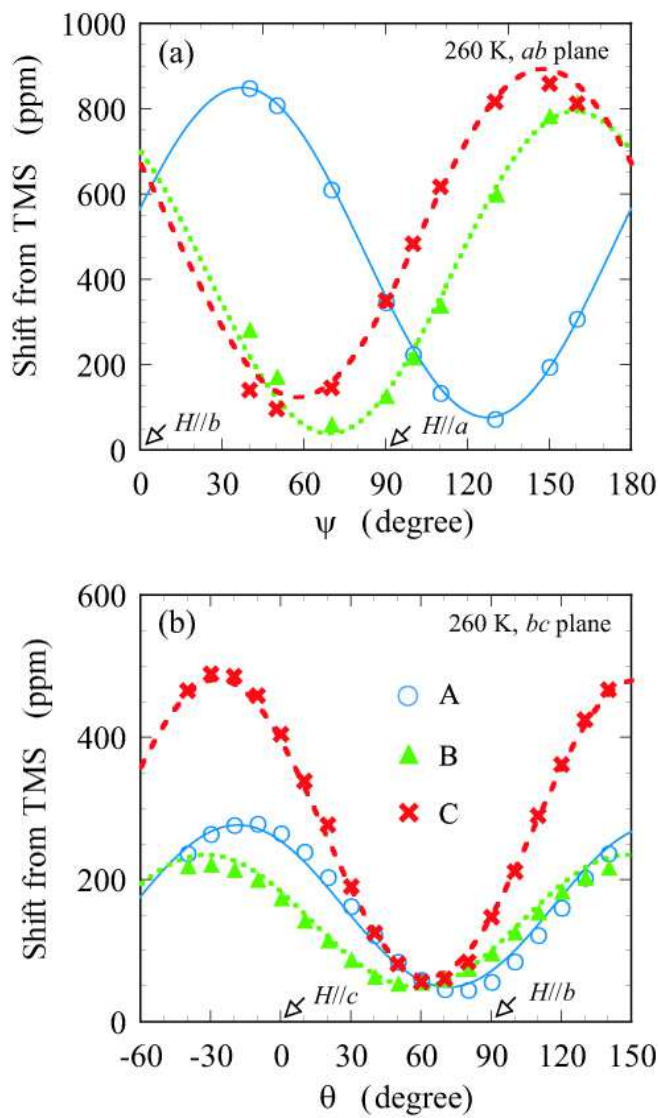}
 \caption{\label{fig4} (Color online) Angular dependence of the $^{13}$C NMR central shift at molecule $i$ (= A, B, and C), $\delta_i$, under (a) $H//ab$ and (b) $H//bc$ at 260 K. Open circles, triangles and crosses stand for the nonequivalent molecules A, B, and C, respectively. The solid, dotted, and dashed lines are fitted curves $\delta_i = [ \delta_i^{xx}(H^x)^2 + \delta_i^{yy}(H^y)^2 + \delta_i^{zz}(H^z)^2 ] / |H|^2$ (see text) with the optimized parameters of $\delta_i^{xx}$, $\delta_i^{yy}$, and $\delta_i^{zz}$ listed in Table~\ref{tab:table1}. }

\end{figure}

   The $^{13}$C NMR shift is proportional to the hyperfine field at the $^{13}$C site projected onto the direction of the external field $H$. 
   The $^{13}$C NMR shift at molecule $i$ (= A, B, and C), $\delta_i$, is defined as the averaged shift of the lines at molecule $i$. 
   In Fig.~\ref{fig4}, we present the representative angular dependences of the $^{13}$C NMR shift $\delta_i$ at 260 K under $H//ab$ [Fig.~\ref{fig4}(a)] and $H//bc$ [Fig.~\ref{fig4}(b)].
   $\delta_i$ is expressed as $\delta_i = [ \delta_i^{xx}(H^x)^2 + \delta_i^{yy}(H^y)^2 + \delta_i^{zz}(H^z)^2 ] / |H|^2$, where $x$, $y$, and $z$ are the ET principal axes [Fig.~\ref{fig3}(a)], $H = (H^x, H^y, H^z)$ is the external field in the ET principal axis [Fig.~\ref{fig3}(b)], and $\delta_i^{xx}$, $\delta_i^{yy}$, and $\delta_i^{zz}$ are the principal components of the $^{13}$C-hyperfine-shift tensor. 
   [Note that the shift tensor for the molecule A with two nonequivalent  $^{13}$C sites is defined as their average, although it can be determined site specifically. 
   The reason why we do so is that the charge and spin densities discussed below in terms of the shift is molecular-specific, and we need the shift value representative of the molecular site. 
   Moreover, this definition is beneficial for the evaluation of electron correlation effects described in Section~\ref{3d}).] 
   We fit the expression to the data in Figs.~\ref{fig4}(a) and ~\ref{fig4}(b), using the crystal structure data given by Sawa \textit{et al}.\cite{Sawa}
   The parameters optimized are listed in Table~\ref{tab:table1}. 
   The fits in Fig.~\ref{fig4} properly capture the sinusoidal angular dependences of $\delta_i$ under both field geometries.
   As seen in Table~\ref{tab:table1}, the shift tensors have different principal values among A, B, and C sites. This evidences differentiations  in local spin susceptibility and charge density among these nonequivalent sites in the unit cell.

\subsection{Temperature dependence of the $^{13}$C NMR shift and the local spin densities}
\label{3c}
\begin{figure*}
 \includegraphics[width=14cm]{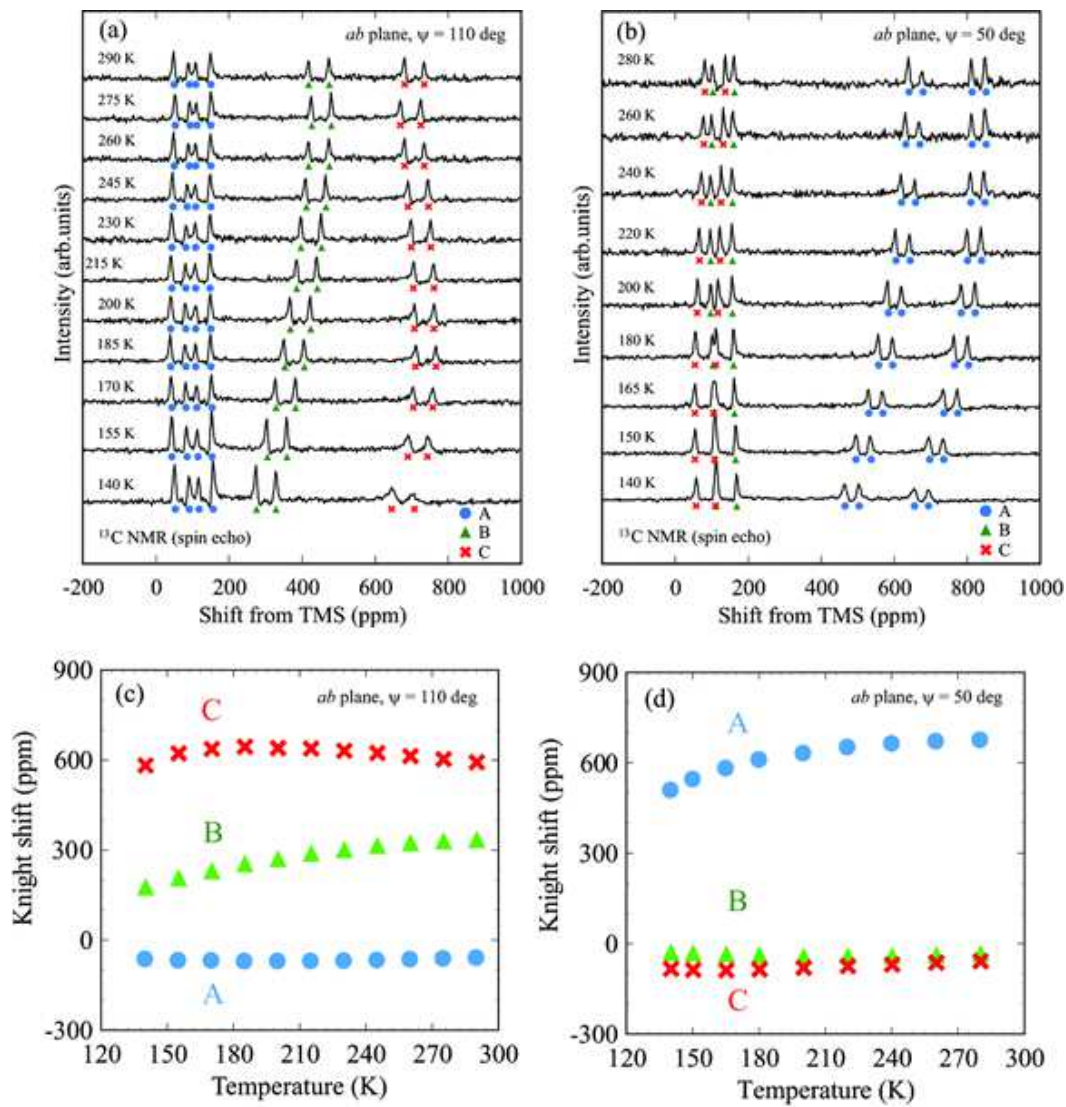}
  \caption{\label{fig5} (Color online) (a), (b) Temperature dependences of the $^{13}$C NMR spectra under $H$ applied within the $ab$ plane ($H//ab$) at (a) $\psi = 110^\circ$ and (b) $\psi = 50^\circ$. [The angle $\psi$ is defined in the inset of Fig.~\ref{fig2}(a).] (c), (d) Corresponding $^{13}$C NMR Knight shifts at A, B, and C nonequivalent molecules.}
\end{figure*}

   Next, we proceed to separate the local charge density and spin density contributions in the observed shift data, and investigate the disproportionation of the local susceptibility within the unit cell and its anomalous temperature dependence.
   Figures~\ref{fig5}(a) and ~\ref{fig5}(b) show the temperature dependences of the $^{13}$C NMR spectra at $\psi = 110^\circ$ and $50^\circ$ (under $H//ab$), respectively. 
   Large temperature dependences are seen in the NMR shift $\delta_i$ when $|\delta_i|$ is large. On the other hand, $\delta_i$ is almost temperature independent when $|\delta_i|$ becomes small.
   The shift $\delta_i$ is the sum of the Knight shift and the chemical shift ($\delta_i = K_i + \sigma_i$), where the Knight shift, $K_i$, is related to the local spin susceptibility $\chi_i$ through $K_i = a_i \chi_i$ and the chemical shift, $\sigma_i$, originates from the orbital motion of electrons within ET molecules.\cite{Klutz} 
   $a_i$ is the hyperfine coupling constant between the $^{13}$C nuclear and itinerant carrier spins. 
   It is well known that the $^{13}$C chemical shift in ET depends on the molecular valence; namely, the amount of hole on ET.\cite{Miyagawa2}
   Because large charge density disproportionations have been observed in $\alpha$-I$_3$ among nonequivalent A, B, and C molecules above $T_\textrm{CO}$,\cite{Kakiuchi, Wojciechowski, Kawai, Moroto, Hirose, Takano} $\sigma_i$ should be determined for each molecule in this compound.  
   Using the principal values of the chemical-shift tensors $\sigma_i^{xx}$, $\sigma_i^{yy}$, and $\sigma_i^{zz}$ (in ppm) determined at 60 K,\cite{Kawai} and linearly interpolating the molecular valence, $\rho_i$, determined by X-ray diffraction\cite{Kakiuchi} to estimate $\rho_i$ at 60 K, we get the following relations between the chemical-shift tensors and valence $\rho_i$; $\sigma_i^{xx} = 130.8 \rho_i + 46.2$, $\sigma_i^{yy} = 30.4 \rho_i + 159.0$, and $\sigma_i^{zz} = -12.1 \rho_i + 61.9$.
   In our case, the temperature dependence of $\sigma_i$ (in ppm) is negligibly small, and the value of $\sigma_i$ is determined as $\sigma_\textrm{A} \sim 165$, $\sigma_\textrm{B} \sim 120$, and $\sigma_\textrm{C} \sim 93$ at $\psi = 110^\circ$, and $\sigma_\textrm{A} \sim 69$, $\sigma_\textrm{B} \sim 165$, and $\sigma_\textrm{C} \sim 166$ at $\psi = 50^\circ$, respectively.
   In general, these chemical-shift tensors allow us to calculate the chemical shift $\sigma_i$ under arbitrary field orientation. 
   Thus, to determine the angular dependence of the Knight shift in the following, we calculated the angular dependences of $\sigma_i$ site-selectively and subtracted them from those of the shift $\delta_i$.
   
\begin{figure}
 \includegraphics[width=8.6cm]{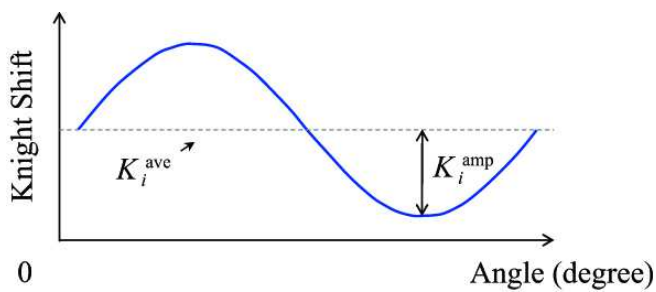}
  \caption{\label{fig6} (Color online) Definitions of the average and amplitude of the angular dependence of the $^{13}$C NMR Knight shift, $K_i^\textrm{ave}$ and $K_i^\textrm{amp}$.}
\end{figure}

   Figures ~\ref{fig5}(c) and ~\ref{fig5}(d) depict the temperature dependences of the NMR Knight shift, $K_i (= \delta_i - \sigma_i)$, under $\psi = 110^\circ$ and $50^\circ$ ($H//ab$), respectively, with the use of the chemical shifts $\sigma_i$ determined above. 
   The Knight shift $K_i$ exhibits strong anisotropy within the $ab$ plane; for instance, $K_\textrm{A}$ is almost zero in Fig.~\ref{fig5}(c), while it is large in Fig.~\ref{fig5}(d). 
   Since the total spin susceptibility, $\chi_\textrm{spin} (\propto 2 \chi_\textrm{A} + \chi_\textrm{B} + \chi_\textrm{C})$, is isotropic in $\alpha$-I$_3$,\cite{Sugano} the anisotropic behavior of $K_i$ should stem from the anisotropy of the hyperfine interaction at the $^{13}$C position, which is expressed as $K_i = [a_i^{xx}(H^x)^2 + a_i^{yy}(H^y)^2 + a_i^{zz}(H^z)^2 ] \chi_i / |H|^2$ with $a_i^{xx}$, $a_i^{yy}$, and $a_i^{zz}$ the principal components of the $^{13}$C hyperfine-coupling tensor [see Fig.~\ref{fig3}(a)].
   [This means that the small value of $K_\textrm{A}$  in Fig.~\ref{fig5}(c) should be attributed to the vanishingly small hyperfine-coupling constant at $\psi = 110^\circ$ geometry.]  
   In ET-based materials, it is known that the profile of the highest occupied molecular orbital (HOMO) of ET determines the anisotropy of the coupling tensor $a_i^{\mu \mu}$ ($\mu$ = $x$, $y$, and $z$).\cite{Kawamoto1}
   The temperature dependence of $a_i^{\mu \mu}$ is negligible, since the spatial distribution of HOMO is most likely to be temperature independent. 
   Moreover, the spatial profile of HOMO is reasonably assumed to be the same at all molecules in the unit cell because the observed differences in the molecular structures are very small.\cite{Sawa}
   Therefore, in the first approximation, we can assume that $a_i^{\mu \mu}$ is site- and temperature-independent.
   We also note that molecules A, B, and C are arranged in a nearly symmetrical manner within the $ab$ plane, as we mentioned above [Fig.~\ref{fig1}(c)].  
   Hence, if $H$ is rotated in the $ab$ plane, the average, $K_i^\textrm{ave}$, and the amplitude, $K_i^\textrm{amp}$, of the angular dependences of the Knight shift [see Fig.~\ref{fig6}] are both expected to reflect the magnitude of $\chi_i$; $K_i^\textrm{ave} = a^\textrm{ave} \chi_i$ and $K_i^\textrm{amp} = a^\textrm{amp} \chi_i$ with $a^\textrm{ave}$ and $a^\textrm{amp}$ the averaged hyperfine-coupling constants.
   
\begin{figure}
 \includegraphics[width=8.6cm]{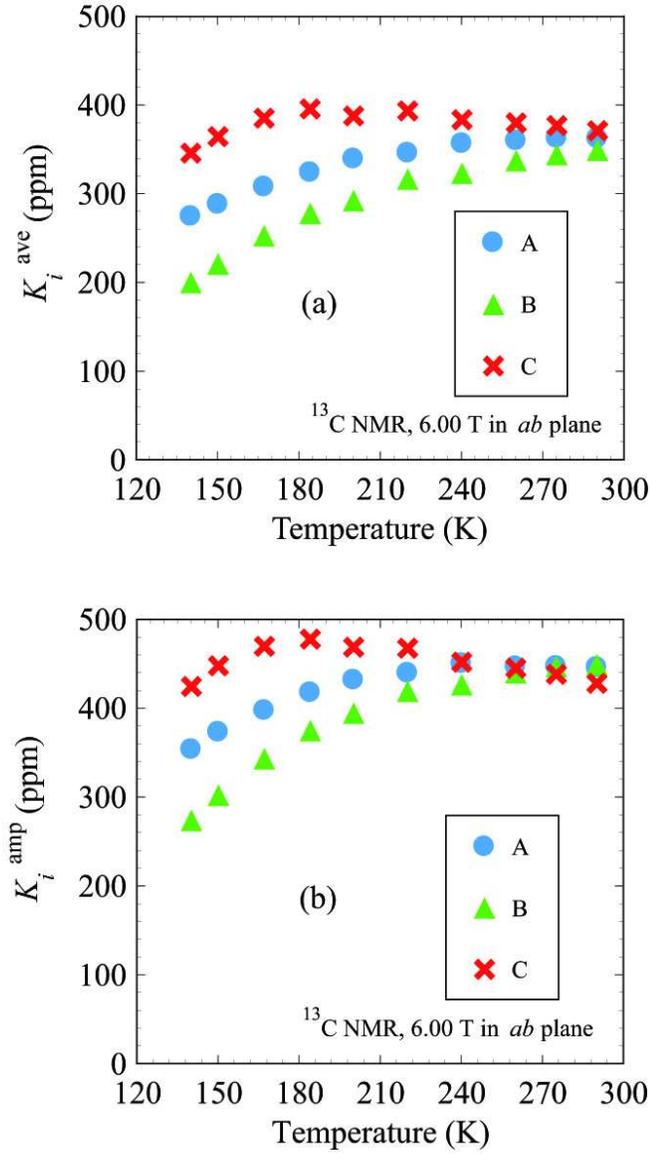}
  \caption{\label{fig7} (Color online) (a) Average ($K_i^\textrm{ave}$) and (b) amplitude ($K_i^\textrm{amp}$) of the sinusoidal angular dependence of the $^{13}$C NMR Knight shift at $i$-site ET molecule ($i$ = A, B, and C) under the external field $H$ applied parallel to the $ab$ plane ($H//ab$).}
\end{figure}

   Figure~\ref{fig7} shows the temperature dependences of $K_i^\textrm{ave}$ and $K_i^\textrm{amp}$.
   For determining them, we calculated the angular dependence of $\sigma_i$, using the empirical relations deduced above, and subtracted them from the angular dependences of the shift, $\delta_i$, at all temperatures measured.
   The external field $H$ was rotated in the $ab$ plane, because $K_i^\textrm{amp}$ becomes largest in this field geometry and the line shapes are readily discerned owing to the anti-phase angular dependence of the lines as mentioned above (in Sec.~\ref{3a}). 
   Around room temperature, there are little site-dependences in the observed $K_i^\textrm{ave}$ and $K_i^\textrm{amp}$. 
   With temperature decreased, however, they exhibit site-specific temperature dependences, and then begin to decrease monotonically below  $T \sim 180$ K at all sites down to $T_\textrm{CO} \sim 135$ K.

\begin{figure}
 \includegraphics[width=8.6cm]{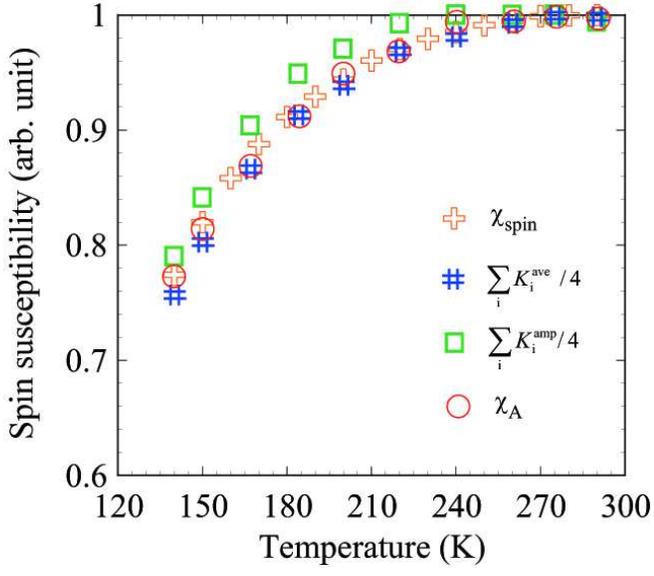}
  \caption{\label{fig8} (Color online) Temperature dependences of the normalized susceptibility $\chi_\textrm{spin}$ (crosses), from Ref.~\onlinecite{Rothaemel}, $\sum_{i} K_i^\textrm{ave}/4$ (sharps), $\sum_{i} K_i^\textrm{amp}/4$ (squares), and $\chi_\textrm{A}$ (circles), respectively. Summations are taken over all molecules in the unit cell.}
\end{figure}

   To deduce the local spin susceptibility $\chi_i$ from $K_i^\textrm{ave}$ ($K_i^\textrm{amp}$), it is necessary to evaluate the hyperfine-coupling constant $a^\textrm{ave}$ ($a^\textrm{amp}$) defined above. 
   This is achieved by comparing $K_i^\textrm{ave}$ ($K_i^\textrm{amp}$) with bulk spin susceptibility $\chi_\textrm{spin}$ at room temperature where the site dependence of  $K_i^\textrm{ave}$ ($K_i^\textrm{amp}$) becomes negligible; that is, $K_i^\textrm{ave} \propto \chi_\textrm{spin}$ ($K_i^\textrm{amp} \propto \chi_\textrm{spin}$).
   At room temperature, $K_i^\textrm{ave} \approx 360$ and $K_i^\textrm{amp} \approx 440$ (in ppm) with little site dependences (Fig.~\ref{fig7}) and $\chi_\textrm{spin} = 6.8 \times 10^{-4}$ emu/mol f.u. measured by Rothaemel \textit{et al}.\cite{Rothaemel} 
   In terms of them, the hyperfine coupling constants are evaluated as $a^\textrm{ave} \approx 5.9$ kOe/$\mu_\textrm{B}$ and $a^\textrm{ave} \approx 7.2$ kOe/$\mu_\textrm{B}$.
   To check whether these coupling constants are applicable under arbitral temperatures, we compare  the temperature dependences of $\sum_{i} K_i^\textrm{ave}/4$ (sharps) and $\sum_{i} K_i^\textrm{amp}/4$ (squares) to that of the spin susceptibility $\chi_\textrm{spin}$ (crosses), as depicted in Fig.~\ref{fig8}, where the summation is taken over all molecules in the unit cell (i.e., molecules A, A, B, and C; see Fig.~\ref{fig1}). 
   All of these quantities are nearly scaled to each other (i.e., $\sum_{i} K_i^\textrm{ave}/4 \propto \chi_\textrm{spin}$ and $\sum_{i} K_i^\textrm{amp}/4 \propto \chi_\textrm{spin}$), which assure our assumptions that $a^\textrm{ave}$ and $a^\textrm{amp}$ are site- and temperature-insensitive.     
   Thus, it is allowed to determine the local susceptibility $\chi_i$ with these coupling constants through $K_i^\textrm{ave} = a^\textrm{ave} \chi_i$ and $K_i^\textrm{amp} = a^\textrm{amp} \chi_i$ at all temperatures measured.
   
   In Fig.~\ref{fig9}(a), we show the temperature dependences of the hereby deduced local spin susceptibility $\chi_i$ from Figs.~\ref{fig7}(a) and ~\ref{fig7}(b). 
   [Note that $\chi_i$ in Fig.~\ref{fig9}(a) is an average among  $K_i^\textrm{ave} / a^\textrm{ave}$ and $K_i^\textrm{amp} / a^\textrm{amp}$ because these values are in good agreements with each other over the entire temperatures measured.] 
   Susceptibility shows little site dependence around room temperature.
   With decreasing temperature, however, large imbalance develops in $\chi_i$ among A, B, and C nonequivalent molecules below $\sim 270$ K, which increases down to $T_\textrm{CO}$.
   Below $\sim 180$ K, all $\chi_i$'s decrease monotonically with decreasing temperature. 
   The tendency, $\chi_\textrm{C} > \chi_\textrm{A} > \chi_\textrm{B}$, is consistent with the preceding works.\cite{Kawai, Hirose, Takano} 
   Just above $T_\textrm{CO}$, its ratio reaches 5:4:6 for A:B:C molecules. 
   This large imbalance cannot be expected by a simple single-band picture.
   Notice that the temperature dependence of $\chi_\textrm{A}$ has the same profile as that of $\chi_\textrm{spin}$; namely, $\chi_\textrm{A} \propto \chi_\textrm{spin}$, as seen in Fig.~\ref{fig8} (circles).
   
   To highlight the temperature dependence of the spin-density disproportionation prominently, we depict the relative local spin density in Fig.~\ref{fig9}(b), defined as $\langle s_i \rangle = \chi_i / [ (2\chi_\textrm{A} + \chi_\textrm{B} + \chi_\textrm{C}) / 4 ]$, which reflects the relative contribution of the $i$-site HOMO to the conduction band around $\varepsilon_F$. 
   Imbalance develops in the relative local spin densities at the B and C sites with decreasing temperature. 
   On the other hand, the A site shows little temperature dependence ($\langle s_\textrm{A} \rangle \approx$ 1), which is the direct consequence of $\chi_\textrm{spin} \propto \chi_\textrm{A}$. 
   At first glance, it seems to be plausible that the observed spin-density disproportionation can be qualitatively understood with the semi-metallic band picture as a redistribution process of B- and C-site HOMO contributions in the two bands. 
   Indeed, band-structure calculations predict two bands locating in the vicinity of the Fermi level $\varepsilon_F$.\cite{Kondo}
   However, it is difficult to explain the strong decrease in $\chi_\textrm{spin}$ (Fig.~\ref{fig8}) by the simple semi-metallic framework; e.g., in the case of 2D Fermi surfaces, $\chi_\textrm{spin} [\propto D(\varepsilon_F)]$ is expected to show little temperature dependence, since the density of states $D(\varepsilon_F)$ is not sensitive to the value of $\varepsilon_F$. 
   Moreover, the trend of the spin-density disproportionation is opposite from what is expected from the x-ray diffraction measurement\cite{Kakiuchi} [B site: (hole) rich, and C site: (hole) poor], which is also anomalous as a conventional semi-metal. 
   As we shall see in Sec.~\ref{4a}, these features should be rather regarded as consequences of the anisotropic Dirac cone realized around $\varepsilon_F$, which was originally predicted under hydrostatic pressures in $\alpha$-I$_3$.\cite{Katayama1}

\subsection{Temperature dependence of the $^{13}$C nuclear spin-lattice relaxation rate and the electron correlation effects}
\label{3d}  
   So far, the local electronic states in $\alpha$-I$_3$ have been revealed in terms of the static susceptibilities. 
   The nuclear spin-lattice relaxation rate, $1/T_1$, probes the fluctuations of electron spins. 
   In this section, we show the site-dependent spin dynamics uncovered by the site-selective measurements of $1/T_1T$, which allows one to see the correlation effects in site-specific or band-specific manners, as discussed in Sec.~\ref{4b}.
   
   In the B and C molecules with inversion center, the two neighboring carbons are equivalent and give the identical relaxation rate. 
   However, the two nonequivalent carbons in the molecule A without inversion center should exhibit different relaxation rates due to different hyperfine coupling constants as reported earlier.\cite{Hennig} 
   Nevertheless, the so-called $T_2$ process, which works to average the site-specific relaxation rates among these two carbons, tends to alter the two values toward some intermediate values in-between them. 
   Then, the distinction of the two observed rates is not so informative, but their average is a meaningful value specific to the A molecule. 
   Thus, we determined the relaxation rate representative of the molecule A from the relaxation curves of the whole spectra (i.e., the quartet), which were nearly single exponential.
   
   In Fig.~\ref{fig10}(a), we present the temperature dependence of the $^{13}$C nuclear spin-lattice relaxation rate, $1/T_1$, divided by temperature $T$ at $\psi = 110^\circ$ under $H//ab$. 
   With decreasing temperature, $1/T_1T$ decreases monotonically at all sites from room temperature down to $T_\textrm{CO}$. 
   The temperature dependence is, however, different from site to site. 
   The decreasing rates are large at A and B sites, while C site exhibits only moderate temperature dependence. 
   In a conventional single-band metal, $1/T_1T$ is not expected to show site-specific temperature dependences because all sites probe the same electronic properties in the conduction band. 
   The distinct behaviors at A, B, and C sites indicate that the electronic structure in this system cannot be interpreted within a simple single-band model. 

   Since the values of $1/T_1TK^2$  measures the degree of electron correlations in the conducting state, $1/T_1TK^2$ was evaluated at each site. 
   However, the Knight shift at A site is too small ( $K_\textrm{A} = -60 \sim -70$ ppm) at this filed geometry in Fig.~\ref{fig5}(c) ($\psi = 110^\circ$) to obtain reliable values of $1/T_1TK^2$, compared to B and C sites. 
   We thus measured the spin-lattice relaxation rate $1/T_1$ on A site at $\psi = 50^\circ$, where the Knight shift at A site is large and hence the relative error in $1/T_1TK^2$ becomes small. 
   The results are shown in Fig.~\ref{fig10}(b). 
   Then, we evaluated the values of $1/T_1TK^2$ for the data series at $\psi = 50^\circ$ for A site, and at $\psi = 110^\circ$ for B and C sites, respectively. 
   In the present case of anisotropic hyperfine couplings, and in the presence of electron correlations, $1/T_1TK^2$ is expressed as the {\itshape modified Korringa relation}; 
$(1/T_{1}T)_i K_i^{-2} = (4{\pi}k_\textrm{B}/\hbar)(\gamma_\textrm{n}/\gamma_\textrm{e})^2\beta_i(\zeta_i, \eta_i)\mathcal{K}_i$  ($i$ = A, B, and C).\cite{Slichter,Kawamoto1} 
   Here, $\gamma_\textrm{e}$ is the gyromagnetic ratio of an electron, $k_\textrm{B}$ is the Boltzmann constant [which yield $(4{\pi}k_\textrm{B}/\hbar)(\gamma_\textrm{n}/\gamma_\textrm{e})^2 = 2.397 \times 10^5$ sec$^{-1}$ K$^{-1}$], $\beta_i(\zeta_i, \eta_i)$ is the $i$-site correction factor for the anisotropy of the hyperfine-coupling tensor as defined in the following ($i$ = A, B, and C)
\begin{widetext}
\begin{center}
\begin{equation}
   \beta_i(\zeta_i, \eta_i) 
= \dfrac{(a_i^{xx} / a_i^{zz} )^2
(\sin^2 \eta_i + \cos^2 \zeta_i \cos^2 \eta_i) 
+ (a_i^{yy} / a_i^{zz})^2
(\sin^2 \eta_i + \cos^2 \zeta_i \cos^2 \eta_i)
+ \sin^2 \zeta_i}
{2 [ (a_i^{xx} / a_i^{zz}) \sin^2 \zeta_i \cos^2 \eta_i
+ (a_i^{yy} / a_i^{zz}) \sin^2 \zeta_i \sin^2 \eta_i
+ \cos^2 \zeta_i ]^2},
\label{eq:two}
\end{equation}
\end{center}
\end{widetext}
and $\mathcal{K}_i$ is the $i$-site {\itshape Korringa ratio}, which reflects the type and strength of the electron correlations.
   The direction of $H$, $(\zeta_i, \eta_i)$ [see Fig.~\ref{fig3}(b)], is determined at $i$ = A, B, and C molecules for the present field geometries, as shown in Table~\ref{tab:table2}, based on the molecular orientations in Ref.~\onlinecite{Sawa}. 
   The principal components of the $^{13}$C hyperfine coupling tensor, $a_i^{xx}$, $a_i^{yy}$, and $a_i^{zz}$, are determined at each molecule $i$ (listed in Table~\ref{tab:table2}) from the total shift tensors, given in Table~\ref{tab:table1}, and the estimated chemical-shift tensors at 260 K (see Sec.~\ref{3c}). 
   The values of $\beta_i(\zeta_i, \eta_i)$ are also presented in Table~\ref{tab:table2}. 

   Figure~\ref{fig11} represents the temperature dependence of the Korringa ratio $\mathcal{K}_i [\propto (1/T_1T)_iK_i^{-2}]$.
   $\mathcal{K}_i = 1$ means that there is no electron correlations. 
   Figure ~\ref{fig11} shows relatively large $\mathcal{K}_i$ for all sites ($\mathcal{K}_i \sim 4 - 11$), indicating the presence of antiferromagnetic correlations.
   At room temperature, $\mathcal{K}_i$ shows little site dependence. 
   Upon cooling, however, it begins to exhibit distinct behaviors for A, B, and C molecules below $\sim 270$ K. 
   The Korringa ratios exhibit only small temperature dependences at A and C sites ($\mathcal{K}_\textrm{A} \approx 7$ and $\mathcal{K}_\textrm{C} \approx 4 - 6$), while $\mathcal{K}_\textrm{B}$ increases below 200 K and reaches a value of 11 at 140 K. 
   In the lowest temperature region, the magnitude relation of $\mathcal{K}_i$ is given by $\mathcal{K}_\textrm{B} > \mathcal{K}_\textrm{A} > \mathcal{K}_\textrm{C}$ . 
   We note that the well-studied dimer-type organic metal $\kappa$-(BEDT-TTF)$_2$Cu[N(CN)$_2$]Br,\cite{Kawamoto1, Kawamoto2, DeSoto} which resides on the verge of the Mott transition, shows $\mathcal{K} \sim 5 - 10$. 
   On the other hand, in $\theta$-(BEDT-TTF)$_2$I$_3$,\cite{Hirata} known as a good 2D metal with a 1/4-filled band, exhibits $\mathcal{K} \approx 2 - 3$, pointing to weak correlations. 
   In our case, $\mathcal{K}_i$ is intermediate or as large as in $\kappa$-type salt. 
   This indicates the importance of electron correlations in $\alpha$-I$_3$, which is discussed in more details in Section~\ref{4b}.

\section{DISCUSSION}
   As we mentioned in the preceding sections (Sections~\ref{3c} and ~\ref{3d}), it is difficult to explain the observed anomalous decreases in spin susceptibility $\chi_\textrm{spin}$ (Fig.~\ref{fig7}), temperature-dependent spin density disproportionations (Fig.~\ref{fig9}), and the large temperature and site dependences of spin fluctuations $\mathcal{K}_i [\propto (1/T_1T)_iK_i^{-2}]$ (Fig.~\ref{fig11}) in terms of simple semi-metallic or single band pictures. 
   Then, how can we understand our NMR results comprehensively? 
   Intriguingly, there are two hints in the previous works. 
   First, as we noted in Sec.~\ref{intro}, resistivity shows weak temperature dependence in this system both at ambient and high pressures.\cite{Schwenk, Tajima2}
   Secondly, the NMR local spin susceptibilities at intermediate\cite{Takano} and high\cite{Hirose} pressures exhibit very close behaviors to our results at ambient pressure. 
   These results suggest that the ambient- and high-pressure conducting states have qualitatively similar features. 
   Furthermore, the observed spin-density disproportionation (Fig.~\ref{fig9}) reflects the presence of site-dependent potentials. 
   The trend of the disproportionation is the same range as that expected in the local site-potentials in Ref.~\onlinecite{Kondo}, which is predicted to stabilize the Dirac cone.\cite{Kondo, Katayama1, Hotta, Mori}
   All of these considerations thus imply that the Dirac cone dominates the electronic properties even at ambient pressure. 
   In fact, as shall be discussed below, the observed features are properly captured by the theoretical consequences of the anisotropic conical dispersion realized around $\varepsilon_F$.\cite{Katayama2} 
   We will focus on the local spin density disproportionations in Sec.~\ref{4a}, and the site-dependent spin fluctuation effects in Sec.~\ref{4b}.

\subsection{Disproportionation of the local spin densities and the site-specific $\bm{k}$-space anisotropy in the conduction band}
\label{4a}
   First, we consider the temperature and site dependences of the local spin densities. 
   Let's assume that the wave function in the conduction band is given by the Bloch sum of the HOMO (highest occupied molecular orbital) at the $i$-site ET molecule ($i$ = A, B, and C), $\phi_i [= \phi(\bm{r} - \bm{R}_l - \bm{\delta}_i)]$, as follows:
\begin{equation}
\Psi_{\bm{k}}(\bm{r}) = \sum_{\bm{R}_l} \sum_{i = \textrm{A}, \textrm{B}, \textrm{C}} 
                        e^{i \bm{k} \cdot \bm{R}_l} C_{i, \bm{k}} \phi (\bm{r} - \bm{R}_l - \bm{\delta}_i),
\end{equation}
where $\bm{R}_l$ stands for the position of one $\alpha$-I$_3$ unit cell, and $\bm{\delta}_i$ is the vector connecting the $i$-site to the A site in the unit cell ($\bm{\delta}_\textrm{A} = 0$). 
   The experimentally obtained $i$-site local spin susceptibility $\chi_i$ is proportional to the thermal average of $\phi_i$ contributing to the conduction band around $\varepsilon_F$, $\bigl\langle \sum_{\bm{k}}|C_{i, \bm{k}}|^2 \bigr\rangle_{\varepsilon = \varepsilon_F \pm k_{\textrm{B}}T}$, which is explicitly given as\cite{Katayama2} $\chi_i = -\int_{-\infty}^\infty d \varepsilon D_i (\varepsilon) f'(\varepsilon) = -\int_{-\infty}^\infty d \varepsilon \sum_{\bm{k}} |C_{i, \bm{k}}|^2 \delta (\varepsilon - \xi_{\bm{k}}) f' (\varepsilon)$, 
with $f(\varepsilon)$ the Fermi-Dirac distribution function, $\xi_{\bm{k}}$ the energy-momentum dispersion of the conduction band, and $D_i(\varepsilon)$ the local electronic density of states at molecule $i$. Prime stands for the derivative with respect to $\varepsilon$. 

   There are two significant consequences of the band-structure calculations relevant to the present results;\cite{Katayama2} one is the large tilting effect of the Dirac cone, and the other is the resulting strong angular dependences of $|C_{i, \bm{k}}|^2$'s about the cone. 
   The ratio of the steepest and gentlest slants (i.e., anisotropy of the Fermi velocities) is estimated at about 10,\cite{Katayama2} and the latter leads to a flat band dispersion giving the van Hove singularity around 10 meV above the Dirac points,\cite{Katayama2} as schematically depicted in Fig.~\ref{fig12}. 
   Noticeably, $|C_{\textrm{B}, \bm{k}}|^2$ is largest around the steepest dispersion, denoted as S in Fig.~\ref{fig11}, and shows a node around the gentle dispersion, denoted as G, whereas $|C_{\textrm{C}, \bm{k}}|^2$ has opposite characteristics.
   $|C_{\textrm{A}, \bm{k}}|^2$ is predicted to show no remarkable angular dependence. 
   This means that the local spin susceptibilities at B and C sites preferentially probe the thermal excitations around the S and G portions of the cone, respectively, while the susceptibility at the A site sees the average over the cone. 
   The overall temperature dependence of the spin susceptibility, shown in Fig.~\ref{fig8}, is basically explained in this context as follows: the conical dispersion has an energy-linear density of states, which gives rise to a linearly temperature-dependent spin susceptibility.\cite{Katayama2, Dora}
   The decreasing total susceptibility with lowering temperature, observed below 200 K in Fig.~\ref{fig8}, can be thought of as signifying this. 
   The leveling-off of the susceptibility at higher temperatures, on the other hand, implies the breakdown of the cone picture at high energies. 
   Actually, the calculated total spin susceptibility based on the band structure tends to saturate at high temperatures.\cite{Katayama2}
   
\begin{figure}
 \includegraphics[width=8.6cm]{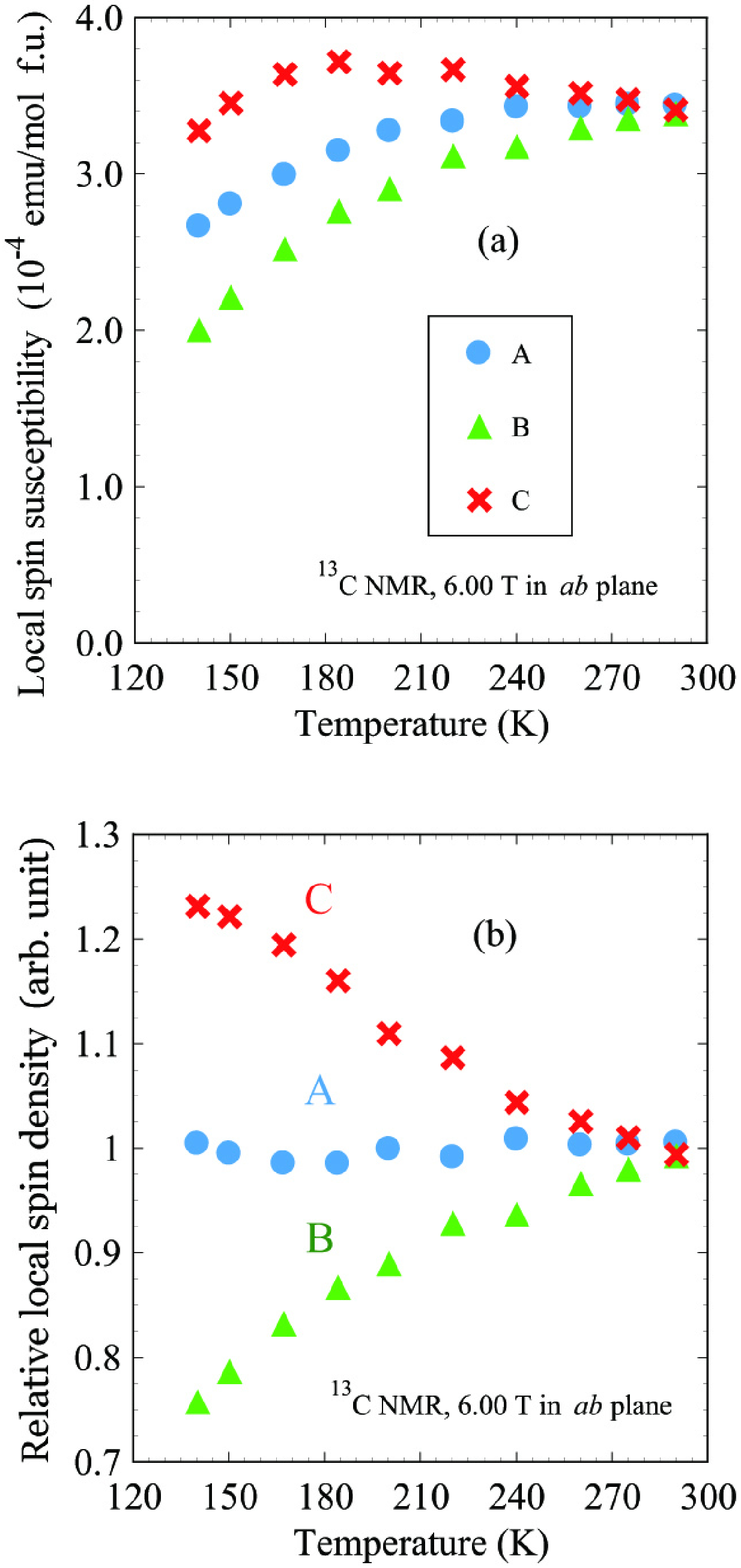}
  \caption{\label{fig9} (Color online) (a) Temperature dependences of the $i$-site local electron spin susceptibility, $\chi_i$ ($i$ = A, B, and C nonequivalent molecules). (b) Relative local spin density, $\langle s_i \rangle = \chi_i / [ (2\chi_\textrm{A} + \chi_\textrm{B} + \chi_\textrm{C}) / 4 ]$.}
\end{figure}

\begin{figure}
 \includegraphics[width=8.6cm]{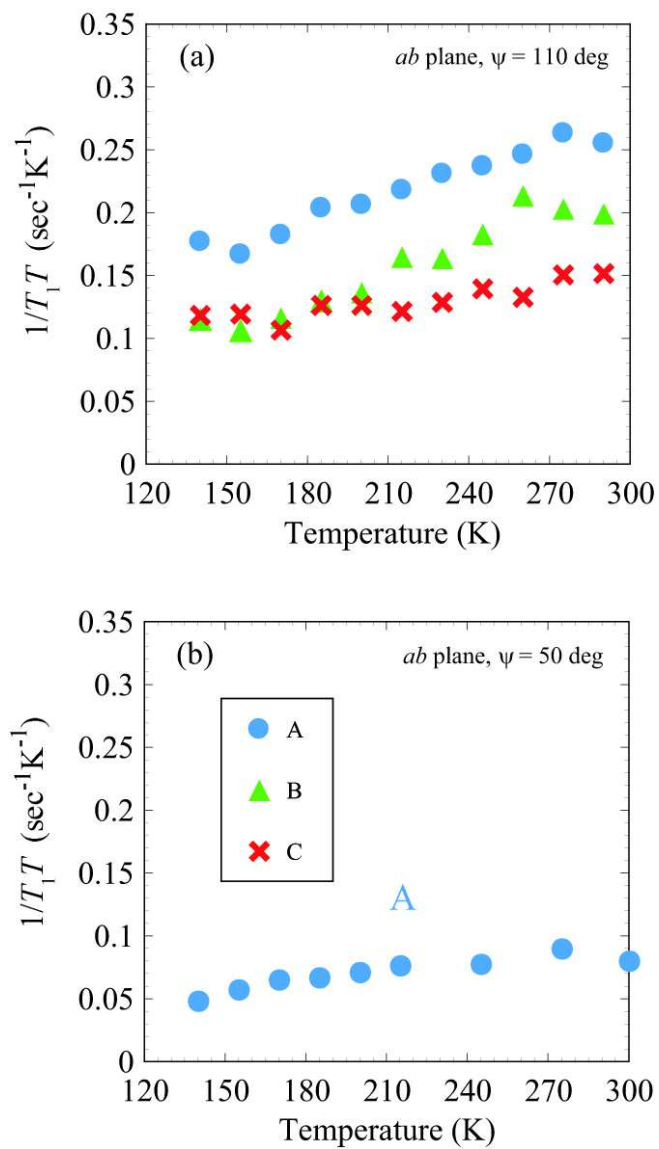}
  \caption{\label{fig10} (Color online) $^{13}$C NMR nuclear spin-lattice relaxation rate divided by temperature, $1/T_1T$, at A, B, and C nonequivalent molecules under (a) $\psi = 110^\circ$ and (b) $\psi = 50^\circ$ ($H//ab$).}
\end{figure}

\begin{figure}
 \includegraphics[width=8.6cm]{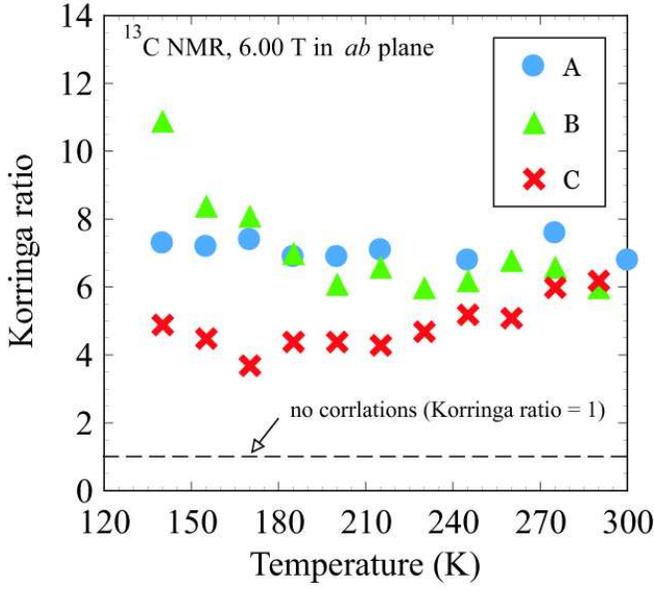}
  \caption{\label{fig11} (Color online) $^{13}$C NMR Korringa ratio $\mathcal{K}_i [\propto (1/T_{1}T)_iK^{-2}_i]$ at $i$ = A, B, and C nonequivalent molecules. Dashed line ($\mathcal{K}_i = 1$) corresponds to the free electron's case. $\mathcal{K}_i > 1$ implies the antiferromagnetic correlations, while $\mathcal{K}_i < 1$ stands for the ferromagnetic ones.}
\end{figure}

\begin{figure}
 \includegraphics[width=8.6cm]{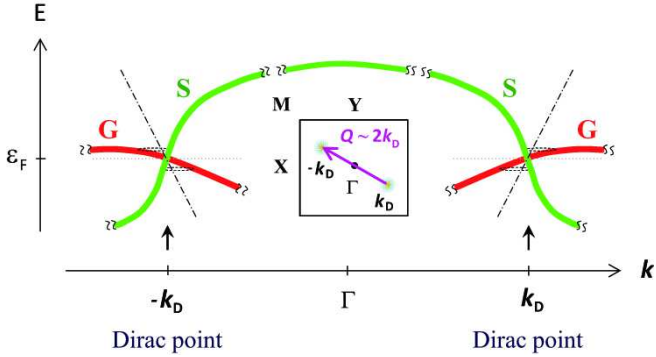}
  \caption{\label{fig12} (Color online) Schematic representation of the $\bm{k}$-space anisotropy of the tilted Dirac cone spectrum predicted by the band-structure calculation given in Ref.~\onlinecite{Katayama2}. S and G stand for the steepest and gentlest portions of the dispersion, and $\bm{k_\textrm{D}}$ ($-\bm{k_\textrm{D}}$) represents the position of the Dirac point. (Inset) First Brillouin zone of $\alpha$-I$_3$. Inter-valley scattering vector with the wave vector $\bm{Q} \sim 2\bm{k_\textrm{D}}$ is represented by the bold arrow.}
\end{figure}

   The spin density disproportionation of $\chi_\textrm{B} < \chi_\textrm{C}$, enhanced at low temperatures [see Figs.~\ref{fig9}(a) and ~\ref{fig9}(b)], reasonably corresponds to the small and large local density of states at the S and G portions on the cone. 
   The peak formation of $\chi_\textrm{C}$ around 190 K is attributable to the van Hove singularity located around G, where $|C_{\textrm{C}, \bm{k}}|^2$ shows the maximum.\cite{Katayama2} 
   On the contrary, $|C_{\textrm{B}, \bm{k}}|^2$ is vanishingly small at G, as mentioned above, which explains why $\chi_\textrm{B}$ continues to increase monotonously up to room temperature. 
   Meanwhile A site shows temperature dependences in between B's and C's both in the local spin susceptibility $\chi_\textrm{A}$ and the relative local spin density $\langle s_\textrm{A} \rangle$ [see Figs.~\ref{fig9}(a) and ~\ref{fig9}(b)]. 
   $\chi_\textrm{A}$ scales to the bulk spin susceptibility, $\chi_\textrm{spin}$, over the whole temperature range ($\chi_\textrm{A} \propto \chi_\textrm{spin}$; see Fig.~\ref{fig7}), which is reflected in the temperature independent profile of $\langle s_\textrm{A} \rangle$ shown in Fig.~\ref{fig9}(b). 
   These results indicate that A site probes the whole conical dispersion in an averaged manner, and supports the theoretical prediction that $|C_{\textrm{A}, \bm{k}}|^2$ shows no remarkable angular dependence on the cone.\cite{Katayama2}
   
   At high temperatures of hundreds Kelvin, the electronic mean free path in molecular conductors can reach the order of the unit cell size due to the strong electron-phonon scatterings. 
   The high-temperature equalization of the local spin susceptibilities [see Fig.~\ref{fig9}(a)] might be partially due to this scattering effect, which will thermally average the anisotropy of the wave functions in the $\bm{k}$-space.
   Several organic conductors exhibit bad metal natures such as a loss in Drude weight in optical conductivity, or broadening in ARPES spectra at high temperatures.\cite{Takenaka, Koizumi} 
   Moreover, the correlation-induced effect such as charge ordering is disturbed at high temperatures in general.
   In the present system, the charge disproportionation, if it is enhanced by electron correlations, is expected to be depressed at high temperatures. 
   This effect tends to make the Dirac point pushed down below $\varepsilon_F$, and may lead to a crossover from the Dirac system to a semi-metal. 
   Actually, the spin susceptibility and Korringa ratio are both temperature- and site-independent around room temperature. 
   It is probable that the picture of the massless Dirac electrons is broken by these thermal effects.

   To illustrate how the hole density at the molecule $i$, $\rho_i$, relates to the above discussed $i$-site spin susceptibility $\chi_i$, we made a comparison between $\rho_i$ and $\chi_i$. 
   Here, the value of $\rho_i$ is estimated from the intra-molecular bond lengths determined by the X-ray diffraction study by Kakiuchi \textit{et al}.\cite{Kakiuchi} and the charge-sensitive modes in the vibrational spectroscopy by Wojciechowski \textit{et al}.\cite{Wojciechowski}
   They found that $\rho_\textrm{B} > \rho_\textrm{C}$, which is supported by Hartree-Fock calculations based on the extended Hubbard model.\cite{Tanaka}
   This relation is, however, opposite to the spin density profile, $\chi_\textrm{B} < \chi_\textrm{C}$. 
   Because our material is a 3/4-filled electron-band (or a 1/4-filled hole-band) system (see Sec.~\ref{intro}), the charge (hole) density might correlate to the spin density, which seems irreconcilable with the present results at first glance. 
   However, the charge-spin correlation holds when both charges and spins are spatially well localized. 
   For itinerant electron systems, on the other hand, they should not necessarily match since the amount of the hole is given by $\rho_i \propto \sum_{\bm{k}} |C_{i, \bm{k}}|^2 [1-f(\xi_{\bm{k}})] \equiv \bigl\langle \sum_{\bm{k}}|C_{i, \bm{k}}|^2 \bigr\rangle_{\varepsilon \ge \varepsilon_F}$
 [with $\sum_{i} \rho_i = 2$ ($i$: all molecules in unit cell)], while the spin density is expressed as $\chi_i \propto \bigl\langle \sum_{\bm{k}}|C_{i, \bm{k}}|^2 \bigr\rangle_{\varepsilon = \varepsilon_F \pm k_{\textrm{B}}T}$ as mentioned above. 
   Obviously, $\chi_i$ is determined by $|C_{i, \bm{k}}|^2$ only in the vicinity of $\varepsilon_F$, while $\rho_i$ reflects the whole summation of $|C_{i, \bm{k}}|^2$ over the conduction band above $\varepsilon_F$. 
   Hence, we acquire the following relations from the experimental results: $\bigl\langle \sum_{\bm{k}}|C_{\textrm{B}, \bm{k}}|^2 \bigr\rangle_{\varepsilon = \varepsilon_F \pm k_{\textrm{B}}T} < \bigl\langle \sum_{\bm{k}}|C_{\textrm{C}, \bm{k}}|^2 \bigr\rangle_{\varepsilon = \varepsilon_F \pm k_{\textrm{B}}T}$ and $\bigl\langle \sum_{\bm{k}}|C_{\textrm{B}, \bm{k}}|^2 \bigr\rangle_{\varepsilon \ge \varepsilon_F} > \bigl\langle \sum_{\bm{k}}|C_{\textrm{C}, \bm{k}}|^2 \bigr\rangle_{\varepsilon \ge \varepsilon_F}$, which suggest that $\sum_{\bm{k}}|C_{\textrm{B}, \bm{k}}|^2 < \sum_{\bm{k}}|C_{\textrm{C}, \bm{k}}|^2$ holds near $\varepsilon_F$ whereas $\sum_{\bm{k}}|C_{\textrm{B}, \bm{k}}|^2 > \sum_{\bm{k}}|C_{\textrm{C}, \bm{k}}|^2$ is valid well above $\varepsilon_F$. 
   These contrasting relations at low and high energies are consistent with the picture of the conical band dispersion which is characterized by the distinct anisotropies in $|C_{\textrm{B}, \bm{k}}|^2$ and $|C_{\textrm{C}, \bm{k}}|^2$;\cite{Katayama2} the portion G (in Fig.~\ref{fig12}), which is contributed largely from the site C, is cut off by the Brillouin zone boundary at low energies, while the portion S relevant to the site B extends to higher energies and over a wide range of the $\bm{k}$-space covering the zone center (see Fig.~\ref{fig12}).

\subsection{Korringa ratios and spin fluctuation effects}
\label{4b} 
   Next, we turn our attentions to the site-specific spin fluctuation effects. 
   As noted in Section~\ref{3d}, the Korringa ratio, $\mathcal{K}_i \propto (1/T_{1}T)_iK^{-2}_i$, is a measure of the strength and type of spin fluctuations in the conduction band.\cite{Moriya}
   As represented in Fig.~\ref{fig11}, $\mathcal{K}_i$ is strongly site and temperature dependent in our system, in contrast to a simple metal where $\mathcal{K}_i$ should be independent on sites and temperatures. 
   This anomalous behavior is understood based on the theoretical consequence that each molecular orbital $\phi_i$ at the site $i$ contributes to the conical dispersion with a differing angular dependence in the $\bm{k}$-space, which is consistent wi         th the discussion in the preceding section. 

   First, it should be reminded that $|C_{\textrm{A}, \bm{k}}|^2$ has a little angular dependence in the reciprocal space.\cite{Katayama2} 
   Hence, $\mathcal{K}_\textrm{A}$ is expected to probe the fluctuations over the entire Dirac cone on average and to serve as a benchmark for the strength of correlations in the system. 
   As we mentioned above, $\mathcal{K}_\textrm{A}$ exhibits a relatively large value ($\mathcal{K}_\textrm{A} \approx 7$) in the whole temperature range (Fig.~\ref{fig11}), pointing the presence of strong or intermediate antiferromagnetic spin fluctuations in the system. 

   The spin-lattice relaxation rate at the site A, $(1/T_1)_\textrm{A}$, is given by\cite{Moriya}
\begin{equation}
(1/T_1)_\textrm{A} = 2 \Bigl( \frac{\gamma_\textrm{n}}{\gamma_\textrm{e}} \Bigr)^2 
\frac{k_\textrm{B}T}{\hbar^2} 
\bigl( a_\textrm{A}^{\perp} \bigr)^2
\sum_{\bm{Q}} \frac{\chi'' (\bm{Q}, \omega)}{\omega},
\end{equation}
where $a_\textrm{A}^{\perp}$ is the transverse component of the hyperfine coupling tensor at the site A, $\omega$ is the NMR resonance frequency, and $\chi'' (\bm{Q}, \omega)$ is the imaginary part of the dynamical spin susceptibility. 
   The wave vector, $\bm{Q}$, and the frequency $\omega$ are characteristic for the spin fluctuations associated with the electron-hole pair excitations at $\varepsilon_F$. 
   For the free electrons, $\chi'' (\bm{Q}, \omega)$ is expressed as $\chi_0'' (\bm{Q}, \omega) = \pi \gamma_\textrm{e}^2 \hbar^3 \omega [D_\textrm{tot}(\varepsilon_F)]/2$,\cite{Moriya} with $D_\textrm{tot}(\varepsilon_F)$ the total density of states at $\varepsilon_{F}$. 
   In the presence of spin fluctuations, on the other hand, $\chi'' (\bm{Q}, \omega)$ are enhanced from the constant value $\chi_0'' (\bm{Q}, \omega)$ at the wave vector $\bm{Q}$ connecting the degenerate states around $\varepsilon_{F}$.
   The present system has two Dirac cones (at valley $\bm{k}_\textrm{D}$ and $-\bm{k}_\textrm{D}$) in the first Brillouin zone, which are mutually connected with time-reversal symmetry.\cite{Katayama1} 
   ($\bm{k}_\textrm{D}$ stands for the position of the Dirac point inside the first Brillouin zone.) 
   In this situation, two kinds of interactions are conceivable: 
(i) the intra-valley scattering with $\bm{Q} \sim 0$, and (ii) the inter-valley scattering with $\bm{Q} \sim 2\bm{k}_\textrm{D}$. 
   The former scattering gives ferromagnetic fluctuations while the latter leads to antiferromagnetic ones. 
   The result of $\mathcal{K}_\textrm{A} \approx 7$ in much excess of unity clearly shows that the inter-valley scattering is the dominant source of the enhanced spin fluctuations in our system. 
   This enhanced electron correlation is naturally understood to help stabilizing the correlation-induced CO-insulating state below 135 K.\cite{Kondo, Hotta}
   
   In contrast to $|C_{\textrm{A}, \bm{k}}|^2$, $|C_{\textrm{B}, \bm{k}}|^2$ and $|C_{\textrm{C}, \bm{k}}|^2$ have so distinct angular and energy dependences on the cone that $\mathcal{K}_\textrm{B}$ and $\mathcal{K}_\textrm{C}$ are expected to measure the fluctuation effects in preferential areas of the $\bm{k}$-space; namely, the S and G portions in Fig.~\ref{fig12}, correspondingly.
   Note that, roughly speaking, $\mathcal{K}_i$ probes the intensity of spin fluctuations per thermally excited quasi-particles instead of the intensity itself because $(1/T_1T)_i$ is divided by $K_i^2$. 
   At low temperatures, $\mathcal{K}_\textrm{B}$ is much enhanced compared with $\mathcal{K}_\textrm{C}$; at 140 K, the value of $\mathcal{K}_\textrm{B}$ reaches 11, which is twice of that of $\mathcal{K}_\textrm{C}$. 
   The thermally excited states probed at B site are well located around the Dirac point in the 2D $\bm{k}$-space because the dispersion of the S portion is steep (see Fig.~\ref{fig12}). 
   In the G portion, on the other hand, the thermal excitation should be extended in a wide range of $\bm{k}$-states because of the flat dispersion with the van Hove singularity located around 150 K above $\varepsilon_F$.
   The different $\bm{k}$-space profile of thermal excitations probed at B and C sites should give a corresponding difference in the inter-valley scattering seen at B and C sites. 
   The inter-valley scattering probed at B site are well restricted to the vicinity of $\bm{Q} \sim 2\bm{k}_\textrm{D}$, whereas that probed at C site should be spread around $\bm{Q} \sim 2\bm{k}_\textrm{D}$.
   The present results imply that the antiferromagnetic fluctuations are sharply enhanced at a wave vector of $\bm{Q} \sim 2\bm{k}_\textrm{D}$, and that the inter-valley scattering is responsible for the electron correlations in this system.
   
   At elevated temperatures, $\mathcal{K}_\textrm{B}$ and $\mathcal{K}_\textrm{C}$ gradually approach each other, and all $\mathcal{K}_i$'s become nearly the same around room temperature ($\mathcal{K}_i \approx 6$). 
   This temperature dependence is understood in line with the diminishing spin density disproportionation at higher temperatures, as discussed in Sec.~\ref{4a} [see Fig.~\ref{fig9}(b)]; the $\bm{k}$-space averaging due to the intense phonon excitations makes the difference between the molecules A, B, and C indistinguishable, and/or the possible depression of charge disproportionation pushes down the Dirac point below $\varepsilon_F$, causing a semi-metallic state.
 
   Theoretically, electron correlations stabilize the Dirac cone dispersion in this system as we mentioned in Sec.~\ref{intro}.\cite{Kondo, Katayama1, Hotta, Mori}
   Furthermore, the presence of the tilted Dirac cone can largely enhance the anisotropy of $|C_{\textrm{B}, \bm{k}}|^2$ and $|C_{\textrm{C}, \bm{k}}|^2$ in the $\bm{k}$-space.\cite{Katayama2}
   The present site-dependent NMR characteristics well below room temperature are consistent with these predictions and give an opportunity to look into the magnetic properties of the tilted Dirac cone in a $\bm{k}$-dependent manner.

\section{CONCLUSION}
   We performed $^{13}$C NMR measurements on the charge-disproportionated conducting state in the layered organic conductor $\alpha$-(BEDT-TTF)$_2$I$_3$ at ambient pressure.
   Reflecting the low crystal symmetry, we obtained separate NMR lines for crystallographically nonequivalent three molecules in the unit cell ($i$ = A, B, and C). 
   The temperature dependences of the resultant site-specific Knight shift $K_i$ are properly captured by the conical dispersion with the Dirac points close to $\varepsilon_F$.
   The analyses of the site-selected nuclear relaxation rate $(1/T_1T)_i$ and $K_i$ point to the presence of strong or intermediate antiferromagnetic spin correlations. 
   Exploiting the theoretical prediction of the peculiar site-specific reciprocal-space anisotropy in the Dirac cone, the present results turn out to show that the local spin susceptibility and electron correlations are both strongly angular dependent on the cone. 
   This outcome is regarded as one of the outstanding aspects inherited from the tilted Dirac cone in $\alpha$-(BEDT-TTF)$_2$I$_3$.

\section{ACKNOWLEDGEMENTS}
   The authors thank R. Kondo and H. Sawa for informing us the structural data and the local electronic density distribution on triiodine anions before publication. 
   They also thank N. Tajima, Y. Suzumura, A. Kobayashi, A. Kawamoto, T. Takahashi, and H. Fukuyama for informing the detailed results of transport experiments at ambient pressure, and valuable and helpful discussions. 
   This work is supported by MEXT Grant-in-Aids for Scientific Research on Innovative Area (New Frontier of Materials Science Opened by Molecular Degrees of Freedom; no. 20110002 and 21110519), JSPS Grant-in-Aids for Scientific Research (A) (no. 20244055) and (C) (no. 20540346), and MEXT Global COE Program at University of Tokyo (Global Center of Excellence for the Physical Sciences Frontier; no. G04).

\begin{table}
\begin{ruledtabular}
\begin{tabular}{cccc} 
$i$ & 
$(\delta_{i})^{xx}$ (ppm) &
$(\delta_{i})^{yy}$ (ppm) &
$(\delta_{i})^{zz}$ (ppm) \\
\colrule
 A & 48 & 77 & 869 \\
 B & 51 & 47 & 780 \\
 C & 59 & 130 & 929 \\
\end{tabular}
\end{ruledtabular}
\caption{\label{tab:table1}
 The $\mu$-axis component of the $^{13}$C-hyperfine-shift tensor $\delta_i^{\mu \mu}$ ($\mu$ = $x$, $y$, $z$; see Fig.~\ref{fig3}) at molecule $i$ ( = A, B, and C) deduced from the angular dependences of the NMR shift $\delta_i$'s at 260 K shown in Figs.~\ref{fig4}(a) and ~\ref{fig4}(b).}
\end{table}

\begin{table*}
\begin{ruledtabular}
\begin{tabular}{ccccccc}
$i$ & 
\shortstack{$a_i^{xx}$ \\ (kOe/$\mu_\textrm{B}$)} & 
\shortstack{$a_i^{yy}$ \\ (kOe/$\mu_\textrm{B}$)} & 
\shortstack{$a_i^{zz}$ \\ (kOe/$\mu_\textrm{B}$)} & 
\shortstack{$\zeta_i$ \\ (degree)} & 
\shortstack{$\eta_i$ \\ (degree)} & 
$\beta_i(\zeta_i, \eta_i)$ \\
\colrule
 A & -1.11 & -1.61 & 13.35 & 21.4 & 57.2 & 0.106 \\
 B & -1.26 & -2.25 & 11.91 & 42.6 & 67.7 & 1.139 \\
 C & -0.74 & -0.70 & 14.32 & 31.8 & 58.8 & 0.281 \\
\end{tabular} 
\end{ruledtabular}
\caption{\label{tab:table2}
Principal components of the hyperfine-coupling tensor $a_i^{\mu \mu}$ ($\mu$ = $x$, $y$, $z$) at molecule $i$, 
spherical polar angles ($\zeta_i$, $\eta_i$) [defined in Fig.~\ref{fig3}(b)], 
and the correction factors for the anisotropy of the hyperfine-coupling tensor $\beta_i(\zeta_i, \eta_i)$ 
[see Eq. (2) in Sec.~\ref{3d}] at $i$ = A ($\psi = 50^\circ$), B ($\psi = 110^\circ$) and C ($\psi = 110^\circ$) molecular sites.}
\end{table*}

\end{document}